# Grid-connected Soft Switching Partial Resonance Inverter for Distributed Generation

Farid Naghavi
Electrical and Computer Engineering Department
Texas A&M University
College Station, TX
farid@tamu.edu

Hamid Toliyat
Electrical and Computer Engineering Department
Texas A&M University
College Station, TX
toliyat@tamu.edu

*Abstract*—This paper presents current control method for a grid-connected partial resonant soft switching inverter. This inverter does not use an electrolytic capacitor and is capable of boosting and bucking the voltage. Grid-connected inverters are used to integrate distributed energy sources to the grid. Current control is vital in meeting the standards and requirements when connecting to the grid. The closed-loop current regulation for this type of converters is analyzed and design guidelines are provided. The control is implemented in the synchronous frame. In addition active damping techniques using capacitor voltage and inductor voltage feedback is used to mitigate CL filter resonance at the output. The mentioned control strategies are implemented on a 400W lab prototype and the results are presented.

## I. Introduction

Distributed generation sources such as wind and solar are connected to the grid thorough a grid connected power electronics converter. These converters inject power to the grid and can also regulated power flow, improve power quality and also control voltage and frequency in case of a grid forming inverter[1]-[2]. With the rising demand for renewable energy sources, distributed generation is increasing as well [2], [4], [14], [20].

Grid-feeding inverters such as Voltage Source Inverter (VSI) and Current Source Inverters (CSI) are controlled to emulate a current source. The references for the active and reactive powers are set by a higher level controller such as a Maximum Power Point Tracking (MPPT) controller [3]. In addition, standards and utility operators have set certain requirements for power quality, voltage ride through, reactive power control [2]. Current controller's role is vital in order to meet these requirements.

The current regulation can be implemented in either the stationary frame or in the synchronous frame. For stationary frame or abc frame current control, proportional resonant (PR) controllers are used in order to achieve zero steady state error. PI controllers in stationary frame suffer from steady state error due to the limited bandwidth of the controller [5]. Alternatively, the current control can be implemented in synchronous frame using a PI controller.

VSIs and CSIs are usually the choice for grid-connected inverters. However, the focus of this work is the partial resonance AC link inverters that were introduced in [6]-[13]. These converters take advantage of partial resonance of an AC link to achieve soft switching. Moreover, large DC electrolytic capacitors are eliminated which, increases the reliability of these converters. This feature makes this converter a suitable candidate in industrial power systems that have high demand for reliability [19].

In [6]-[10] the parallel AC link inverters have been proposed. The AC link consists of an inductor in parallel with a small film capacitor. To add galvanic isolation, the magnetizing inductance of a transformer can be utilized as well. Therefore, in the isolated version [6]-[7] the AC link consist of the magnetizing inductance of transformer and two small film capacitors on the primary and secondary of the transformer. In [11]-[12] the series version of the converter is presented. Modular converters with stacked cells have also been presented in [9]-[10], [12]. While these references were focused on topology and operation principles of such converters, the focus of this work is the current regulation and closed-loop control in a grid-connected, grid-feeding inverter.

This work focuses on the isolated parallel partial resonance DC-AC converter as shown in Fig. 1. This converter is capable of both buck and boost operation. In addition, soft switching for all the switches of the converter is achieved at all conditions and operating points. Besides, high dv/dt transitions are not present in this inverter. These features make this converter a suitable candidate for integration of distributed generation sources to the grid. One drawback of this inverter is the variable switching frequency that is dependent on the power. If the power is increases the switching frequency is decreased. Consequently, it is essential to consider the lowest switching frequency when designing the control loops.

In order to mitigate the switching frequency harmonics a CL filter is used on the output side of the converter. Due to the low damping in the CL filter, resonance oscillations can happen. The CL filter can also make the current loop unstable. In order, to suppress the oscillations an active damping method using capacitor voltage feedback is implemented. The current control and the active damping loops are implemented in the synchronous frame.

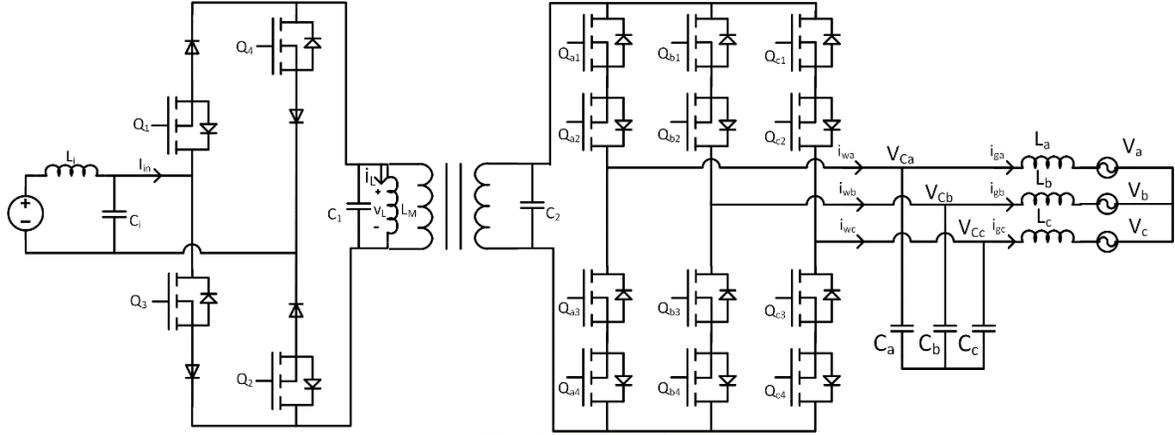
Fig. 1 Converter topology

The rest of this paper is organized as follows, section II looks at the topology and operation principles of the inverter. Section III details the control strategy and role of each control loop. Section IV provides guidelines for parameter selection, section V presents the results and section VI concludes the paper.

## II. OPERATION PRINCIPLES

### A. Topology

Fig. 1, shows the partial resonance AC link inverter. The converter utilizes bidirectional reverse blocking switches which enables bidirectional power flow. The AC link consist of the magnetizing inductance of the transformer ($L_M$) in parallel with two small film capacitors ($C_1$ and $C_2$). The power transfer takes place indirectly between the input and the output through the magnetizing inductance. The link is charged from the input first, after short partial resonance periods, the energy is discharged to the output. The partial resonance periods enable soft transitions which enable the soft switching operation of the converter.

Since the energy is stored in the inductor, during discharge modes, the converter output switches can be controlled to shape the output currents according to the current references. This feature gives the converter some resemblance to CSIs, however, this converter does not have a stiff current in contrast to a CSI.

### B. Operation Modes

The magnetizing inductance ($L_M$) is charged from the input and then discharged to the output. The output currents are shaped according to their references. The converter has 12 operation modes that are briefly discussed in this section. Among these modes, two modes are charging modes, 4 modes are discharging and 6 modes are partial resonance modes. The link cycle for a few switching cycles is shown in Fig. 2.

*Mode 1:* In mode 1, the magnetizing inductance is charged from the input. The input current reference is generated by the higher level controller according to power or current loops. The link voltage has peaked in previous mode and is descending during mode 1. When the link voltage is equal to the input voltage, the input switches $Q_1$ and $Q_2$ become forward biased and turn on. The turn-on of $Q_1$ and $Q_2$ is with Zero Voltage Switching (ZVS) since the input and link voltage ($V_L$) are equal.

The input switches start conducting until the average input current ($I_{in}$) is equal to the input reference current. This method of control is also known as charge control. During this mode the link voltage is constant and the link current increase and the link is charged. When the input reference current is met the switches are turned off. Due to the presence of the input capacitor ($C_i$) and link capacitor ($C_1$), during turn-off the voltage across the switches are nearly zero. As a result turn-off of the switches happen with ZVS as well.

*Mode 2:* All of the switches are off during this mode. Therefore, the magnetizing inductance ($L_M$) and link capacitors ($C_1$ and $C_2$) start to resonate. This mode is a partial resonance mode. The link voltage decreases until it becomes equal to the line-line voltage of the phase selected for mode 3.

*Mode 3:* The stored energy in the link is discharge to two output phase pairs that are selected for this mode and mode 5. These phase pairs are selected based on their current references.

For example, if the reference for phase A is positive, switch $Q_{a3}$ is turned on and if it's negative $Q_{a1}$ is turned on. Therefore, three switches from three legs of the output side are selected. At the beginning of mode 3, all of these switches are reverse biased so as the converter enters mode 3, all these gate signals are generated. The link voltage decreases until the switches of one phase pair become forward biased e.g. A and B. At this point, the switches start conducting and the link starts to discharge. This mode continues until the average output current ($I_{wx}$) for one of the phases meets its reference e.g. phase B. The switch for that phase is turned off and the link starts another partial resonance. Similar to mode 1, the switch turn-on and turn-off for the output phases are also ZVS with the same mechanism.

*Mode 4:* This mode is another partial resonance mode. The link voltage continues to decrease until it becomes equal to the voltage of the next phase pair for mode 5. Although the gate signals for two of switches in this mode are still on, the switches do not conduct as they are reverse biased.

*Mode 5:* When the link voltage is equal to the other phase pair e.g. A and C, the switches start conducting. This mode continues until there's a small residual energy left in the link. At this point the other phase pair e.g. A and C should also meet their references. This is due to the fact that higher control loops ensure power balance between input and output of the converter. The

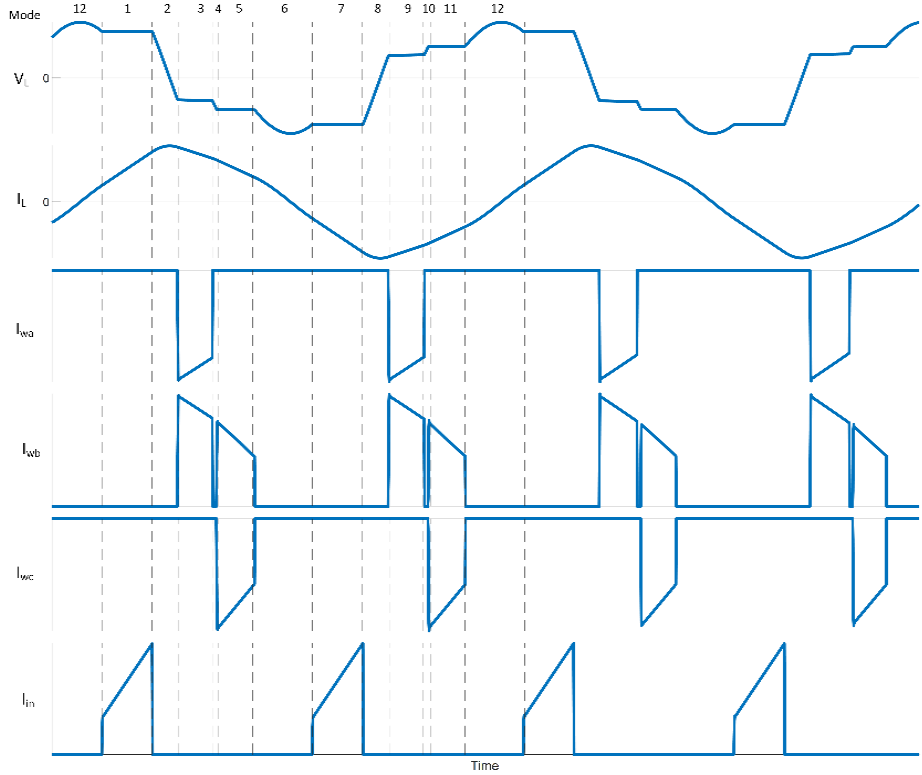
Fig. 2 Link cycles and operation modes

residual energy should be enough to allow the link voltage to decrease to its negative peak in mode 6.

*Mode 6:* Because of the residual energy that was left in the link in mode 5, the link continues to resonate. The link reaches its negative peak in this mode and starts to increase.

*Modes 7-12:* These modes are similar to modes 1-6 with the difference that the link charges and discharges with opposite polarities.

It can be observed that all of the switches in this converter are soft switched with the same mechanism described above. In fact, the soft switching is maintained throughout all the operating points of the inverter.

### C. Switch Controller

Similar to PWM converters, the switch controller acts as a modulator. The switch controller, decides when to turn on or turn off the switches based on the sequence of modes that was explained in section II.B. Fig. 3, shows the switch controller. The switch controller can be modeled as a delay, $G_d(s)$, equal to half the switching period because the reference are updated twice per switching cycle.

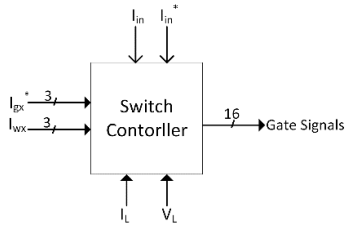
Fig. 3 Switch Controller

## III. CONTROL STRATEGY

There are multiple loops in the overall control of the converter. Fig. 4, shows the control block diagram. It consists

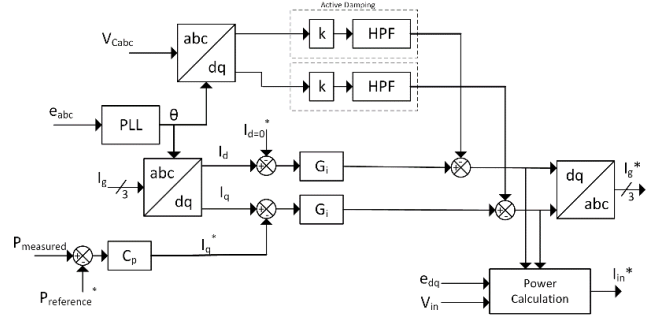
Fig. 4 Control diagram

of active damping loop, outer current control loop and the power control loop.

### A. Active Damping Loop

Active damping or virtual impedance loop is the inner most loop. Due to the CL filter at the output of the converter, resonance oscillations happen. In addition, it can make the current control loop unstable.

The active damping or virtual impedance technique, modifies the current reference of the inverter to emulate an impedance in the filter without physically adding one. There are several methods to implement the active damping. It can be implemented either in the synchronous frame or in the stationary frame. Capacitor voltage or inductor feedback have been reported in the literature for a CL filter in current source converters [15]-[18].

Capacitor voltage or the inductor voltage signal in the synchronous frame can be used for the active damping of the filter. The capacitor voltage is measured and then transformed to the dq axes voltages. The damping coefficient is selected based on the converter parameters. A high pass filter (HPF) is also used to filter out the DC component of the dq voltages in order to prevent reference saturation. The capacitor dq voltages in the synchronous frame have both DC and AC components. The grid frequency component of the capacitor voltage is the DC component and the other harmonics in the capacitor voltage is the AC component.

Therefore, the active damping loop blocks the DC voltages and passes the AC components including the CL filter's resonant harmonics. These AC components are subtracted from the current reference to the inverter.

### B. Current Loop

As shown in Fig. 4, the current is regulated in the synchronous frame using PI controllers. With the transformation convention used in this work, $I_q$ controls the active power and $I_d$ controls the reactive power. The references for the dq axes currents are generated by higher level controllers such as MPPT. The reactive power reference for the inverter is set to zero ($I_d=0$) for unity power factor operation.

### C. Grid Synchronization

A phase-locked loop (PLL) is used for grid synchronization. The PLL block provides the angle for abc frame to dq frame transformation.

## IV. PARAMETER SELECTION

### A. Active Damping using Capacitor Voltage Feedback

Fig. 5(a), shows the control block diagram of the active damping loop. Fig. 5(b) shows the simplified block diagram.

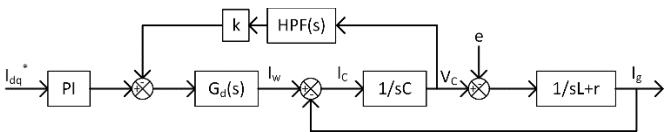

Fig. 5(a) Active damping control block diagram

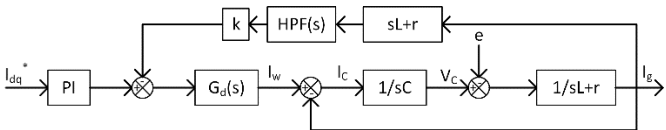

Fig. 5(b) Simplified block diagram

The delay associated with the converter is shown as $G_d(s)$. The delay is equal to half a switching period since the references are updated twice per switching cycle as described in section II. A Pade expansion is used to model the delay

$$G_d(s) = e^{-sT_d} = \frac{\frac{1}{12}T_d^2s^2 - \frac{1}{2}T_Ls+1}{\frac{1}{12}T_d^2s^2 + \frac{1}{2}T_ds+1} \quad (1)$$

Multisampling is used for the control loops therefore, the calculation delay is not considered in the delay calculation. The forward loop gain for the active damping loop can be calculated as

$$G_{AD}(s) = G_d(s)A(s)G_p(s) \quad (2)$$
$$A(s) = k(sL + r)HPF(s) \quad (3)$$
$$HPF(s) = \frac{s}{s+2\pi f_c} \quad (4)$$
$$G_p = \frac{1}{LCs^2+rCs+1} \quad (5)$$

Fig. 6 shows the Bode plot of the grid current to the inverter current transfer function without active damping (blue) for the converter listed in Table I. Clearly it becomes unstable at the resonant frequency. The switching frequency is 6 kHz as the worst case scenario.

The active damping forward loop gain should be stable and provide enough gain at harmonic frequencies. A rule of thumb is to set the bandwidth to 1/10th of the switching frequency. Due to the added term (sL+r) in A(s), the corner frequency of the HPF should be close to the zero created by (sL+r). Since the added term reduces the gain significantly the corner frequency of the HPF should be placed close to this added zero to increase the gain. The zero is usually close to the origin ($\frac{r}{L}$).

One drawback with the capacitor voltage damping is that the corner frequency should be small to compensate for the gain. As a result is leads to slow dynamic. For the converter of Table I the zero is at 8Hz and the HPF corner frequency is set to 320Hz. The damping coefficient k is set to 0.2 to achieve a stable and damped response. At higher corner frequencies the active damping loop become less effective.

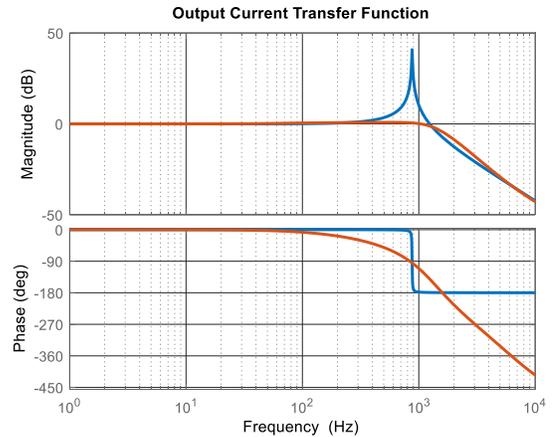

Fig. 6 Grid current Bode plots without active damping (blue) and with active damping (red)

### B. Active Damping using Inductor Voltage Feedback

Another alternative for the feedback signal of the active damping is the inductor voltage feedback. Since both the grid voltage and capacitor voltage are measured and transformed in the dq reference frame quantities, the inductor voltage can be calculated easily by subtracting the capacitor voltage from the grid voltage. Since the DC component of the capacitor voltage and inductor voltage in the synchronous frame are close to each other, the high pass filter can be removed. This will result in a faster active damping response. Fig. 7 shows the block diagram of the active damping using inductor voltage feedback. The frequency response is the same as the one shown in Fig. 6 and the damping factor (k) can be selected in a similar fashion. Using the inductor voltage feedback also increases disturbance rejection from the grid voltage.

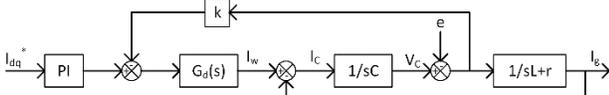

Fig. 7 Active damping block diagram using inductor voltage feedback

## C. Current Loop

The current loop bandwidth is set to below 1/10th of the active damping loop to prevent interference between these two loops.

## RESULTS

A 400W inverter with the specifications in Table I is designed and will be used to implement the different control loops to verify the proposed control algorithms.

TABLE I INVERTER PARAMETERS

| | |
|---|---|
| Input Voltage (V) | 150 |
| Output Voltage (V) | 60 |
| Output Current (A) | 1-4.5 |
| Output Filter Inductance (mH) | 1 |
| Output Filter Capacitance (µF) | 33 |
| Damping Factor, k | 0.2 |
| HPF cut-off frequency (Hz) | 160 |
| $K_i$ | 800 |
| $K_p$ | 0.2 |
| Magnetizing Indurance(µH) | 425 |
| Turns ratio | 1:1 |
| Link Capacitance(nH) | 100 |
| Switching Frequency(kHz) | 6-10 |

Fig. 8 shows the grid current during start up. As seen due to the slow dynamic of the HPF there's some oscillations at the beginning. The THD at steady state is 3.1%.

Fig. 9, shows the grid voltage and current of phase A. The current and voltage are in phase therefore, the current loop is able to control the power factor. Fig. 10 and 11 show the grid current during transients with q-axis current changing from 2A to 4A. Fig. 11 shows the grid currents in the dq synchronous frame. As seen, during transient the slow dynamic of the HPF causes overshoot in d-axis. The THD at steady state reaches 3.1% for these conditions.

Figs. 12 and 13 show the grid currents when inductor voltage feedback is used for active damping. By comparing Figs 12 and 10, it's evident that removal of the HPF from the active damping loop improves the dynamic response during transients. Additionally, the THD at steady state is 1.7% which is lower than capacitor voltage feedback. Fig. 13, shows that the dynamics of the $I_q$ and $I_d$ is improved as well with use of inductor voltage feedback.

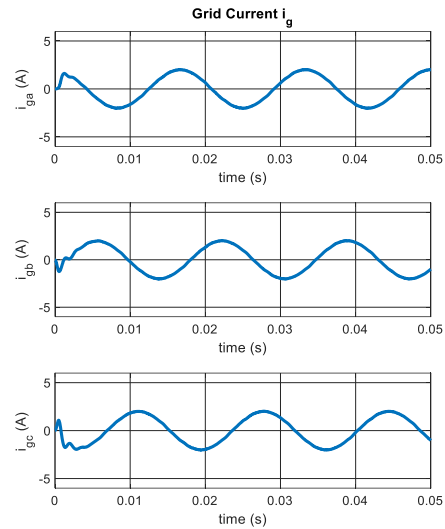

Fig 8 Grid currents in abc frame with capacitor voltage feedback

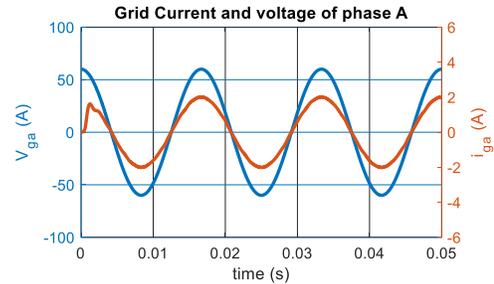

Fig. 9 Grid current and voltage of phase A with capacitor voltage feedback

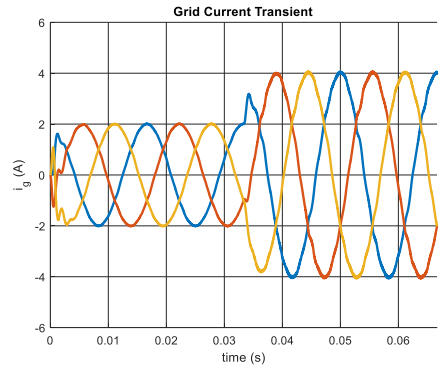

Fig. 10 Grid current during reference change from Iq=2A to Iq=4A

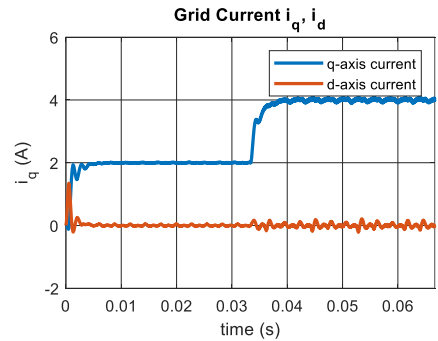

Fig. 11 $I_q$ and $I_d$ during transients with capacitor voltage feedback

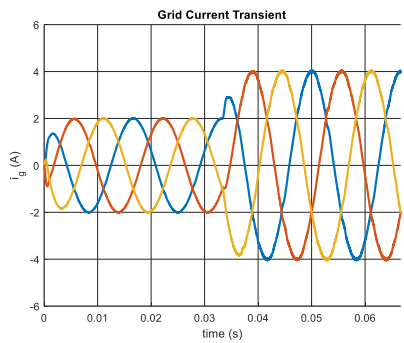

Fig. 12 Grid current transients in abc frame with inductor voltage feedback

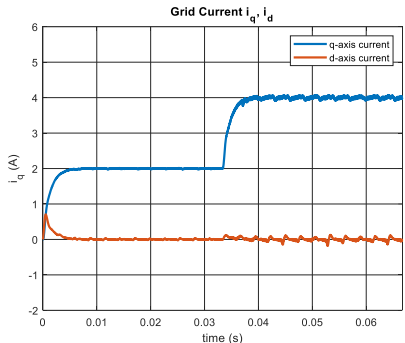

Fig. 13 $I_q$ and $I_d$ during transients with inductor voltage feedback

## V. CONCLUSION

This article proposes control methods aimed at the partial resonance AC link inverters. These inverters take advantage of partial resonance of an AC link to achieve soft switching for all the switches. Soft switching conditions are maintained for all the operation points of the converter. Moreover, due to the lack of electrolytic capacitors these converters have high reliability. However, these converters have variable switching frequency. As the power increase the switching frequency decreases. As a results, the control loops should be designed for the minimum switching frequency. Also, the transformer operation is similar to a flyback transformer or a coupled inductor.

The proposed current control is implemented in the synchronous frame. Due to the use of CL filters at the output of the converter, active damping techniques should be used to suppress the resonance oscillations of the filter. Both capacitor voltage and inductor voltage feedback signals were used for the active damping loop. Inductor voltage feedback has a better dynamic performance because of absence of the HPF. The control strategy for each loop was discussed and design guidelines and parameter selection were provided. The results were presented that showed the performance of the adopted control strategy.

REFERENCES

[1] J. M. Guerrero, P. C. Loh, T. Lee and M. Chandorkar, "Advanced Control Architectures for Intelligent Microgrids—Part II: Power Quality, Energy Storage, and AC/DC Microgrids," in *IEEE Transactions on Industrial Electronics*, vol. 60, no. 4, pp. 1263-1270, April 2013

[2] A. Timbus, M. Liserre, R. Teodorescu, P. Rodriguez and F. Blaabjerg, "Evaluation of Current Controllers for Distributed Power Generation Systems," in *IEEE Transactions on Power Electronics*, vol. 24, no. 3, pp. 654-664, March 2009

[3] J. Rocabert, A. Luna, F. Blaabjerg and P. Rodríguez, "Control of Power Converters in AC Microgrids," in *IEEE Transactions on Power Electronics*, vol. 27, no. 11, pp. 4734-4749, Nov. 2012

[4] R. Rahimi, S. Habibi, P. Shamsi and M. Ferdowsi, "A Three-Winding Coupled Inductor-Based Dual-Switch High Step-Up DC-DC Converter for Photovoltaic Systems," in *IEEE Journal of Emerging and Selected Topics in Industrial Electronics*.

[5] D. G. Holmes, T. A. Lipo, B. P. McGrath and W. Y. Kong, "Optimized Design of Stationary Frame Three Phase AC Current Regulators," in *IEEE Transactions on Power Electronics*, vol. 24, no. 11, pp. 2417-2426, Nov. 2009

[6] H. Keyhani and H. A. Toliyat, "Single-Stage Multistring PV Inverter With an Isolated High-Frequency Link and Soft-Switching Operation," in *IEEE Transactions on Power Electronics*, vol. 29, no. 8, pp. 3919-3929, Aug. 2014

[7] H. Keyhani, M. Johnson and H. A. Toliyat, "A soft-switched highly reliable grid-tied inverter for PV applications," *2014 IEEE Applied Power Electronics Conference and Exposition - APEC 2014*, 2014,

[8] M. Amirabadi, A. Balakrishnan, H. A. Toliyat and W. C. Alexander, "High-Frequency AC-Link PV Inverter," in *IEEE Transactions on Industrial Electronics*, vol. 61, no. 1, pp. 281-291, Jan. 2014

[9] M. Moosavi, F. Naghavi and H. A. Toliyat, "A Scalable Soft-Switching Photovoltaic Inverter With Full-Range ZVS and Galvanic Isolation," in *IEEE Transactions on Industry Applications*, vol. 56, no. 4, pp. 3919-3931, July-Aug. 2020

[10] Y. M. Moosavi and H. A. Toliyat, "A Multicell Cascaded High-Frequency Link Inverter With Soft Switching and Isolation," in *IEEE Transactions on Industrial Electronics*, vol. 66, no. 4, pp. 2518-2528, April 2019

[11] M. Khodabandeh, B. Lehman and M. Amirabadi, "A Highly Reliable Single-Phase AC to Three-Phase AC Converter With a Small Link Capacitor," in *IEEE Transactions on Power Electronics*, vol. 36, no. 9, pp. 10051-10064, Sept. 2021

[12] E. Afshari and M. Amirabadi, "An Input-Series Output-Parallel Modular Three-Phase AC–AC Capacitive-Link Power Converter," in *IEEE Transactions on Power Electronics*, vol. 36, no. 12, pp. 13603-13620, Dec. 2021

[13] F. Naghavi, H. Toliyat, "A Soft-Switching Single-Stage AC-DC Converter," *2022 IEEE Texas Power and Energy Conference (TPEC)*, 2022

[14] A. P. Sirat, H. Mehdipourpicha, N. Zendehdel and H. Mozafari, "Sizing and Allocation of Distributed Energy Resources for Loss Reduction using Heuristic Algorithms," *2020 IEEE Power and Energy Conference at Illinois (PECI)*, 2020

[15] Y. W. Li, "Control and Resonance Damping of Voltage-Source and Current-Source Converters With LC Filters," in *IEEE Transactions on Industrial Electronics*, vol. 56, no. 5, pp. 1511-1521, May 2009

[16] F. Liu, B. Wu, N. R. Zargari and M. Pande, "An Active Damping Method Using Inductor-Current Feedback Control for High-Power PWM Current-Source Rectifier," in *IEEE Transactions on Power Electronics*, vol. 26, no. 9, pp. 2580-2587, Sept. 2011

[17] Z. Bai, H. Ma, D. Xu, B. Wu, Y. Fang and Y. Yao, "Resonance Damping and Harmonic Suppression for Grid-Connected Current-Source Converter," in *IEEE Transactions on Industrial Electronics*, vol. 61, no. 7, pp. 3146-3154, July 2014

[18] X. Wang, Y. W. Li, F. Blaabjerg and P. C. Loh, "Virtual-Impedance-Based Control for Voltage-Source and Current-Source Converters," in *IEEE Transactions on Power Electronics*, vol. 30, no. 12, pp. 7019-7037, Dec. 2015

[19] N. Torabi, F. Naghavi and H. A. Toliyat, "Real-time fault isolation in multiphase multilevel NPC converters using active semi-supervised fuzzy clustering algorithm with pairwise constraints," *2017 IEEE International Electric Machines and Drives Conference (IEMDC)*, 2017

[20] A. Shojaeighadikolaei, A. Ghasemi, A. G. Bardas, R. Ahmadi and M. Hashemi, "Weather-Aware Data-Driven Microgrid Energy Management Using Deep Reinforcement Learning," *2021 North American Power Symposium (NAPS)*, 2021